\documentclass[10pt,aps,prb,twocolumn,amsmath,notitlepage,floatfix,footinbib,superscriptaddress,showpacs, showkeys]{revtex4-1}
\usepackage{graphicx}
\usepackage{amssymb}
\usepackage[usenames,dvipsnames]{color}
\usepackage{hyperref}
\begin{document}
\title{Currents and Green's functions of impurities out of equilibrium -- results from inchworm Quantum Monte Carlo} 

\author{Andrey~E.~Antipov}
\author{Qiaoyuan Dong}
\author{Joseph Kleinhenz}
\affiliation{Department of Physics, University of Michigan, Ann Arbor, Michigan 48109, USA}
\author{Guy Cohen}
\affiliation{School of Chemistry, Tel Aviv University, Tel Aviv 69978, Israel}
\author{Emanuel Gull}
\affiliation{Department of Physics, University of Michigan, Ann Arbor, Michigan 48109, USA}

\date{\today}

\begin{abstract} 
We generalize the recently developed inchworm quantum Monte Carlo method to the full Keldysh contour with forward, backward, and equilibrium branches to describe the dynamics of strongly correlated impurity problems with time dependent parameters. We introduce a method to compute Green's functions, spectral functions, and currents for inchworm Monte Carlo and show how systematic error assessments in real time can be obtained. We then illustrate the  capabilities of the algorithm with a study of the behavior of quantum impurities after an instantaneous voltage quench from a thermal equilibrium state.
\end{abstract}

\maketitle

\section{Introduction}

The dynamical response of strongly correlated electron systems exhibits a fascinating interplay between quantum mechanics and dissipative statistical mechanics. The description of such dynamics is challenging, as the presence of electronic correlations requires a non-perturbative description beyond what is possible with standard analytical tools.

The Anderson impurity model\cite{Anderson61} with time-dependent parameters or multiple baths at different thermodynamic parameters\cite{Meir93} is one of the simplest fermionic non-equilibrium quantum problems. It contains localized impurity states coupled to a non-interacting bath and appears in a wide range of contexts, including impurities embedded into a host material,\cite{Anderson61} confined nanostructures,\cite{Hanson07} and molecules adsorbed on surfaces.\cite{Brako89,Langreth91} 

Impurity models also appear as auxiliary models in non-equilibrium dynamical mean field theory.\cite{Aoki14,Schmidt02,Freericks06} A solution of the dynamical mean field equations requires the calculation of two-time quantities such as the spectral function, for temperatures low enough that the impurity model exhibits Kondo behavior and---in quench setups---for times long enough that a new equilibrium or steady state behavior can be observed.\cite{Meir93} Additionally, as intuition gained in equilibrium may not be accurate in correlated time-dependent situations, methods that either are numerically exact or offer a stringent assessment of uncertainties are desired.

In this paper, we present a numerical method that fulfills all of these criteria. It can treat general time-dependent setups of Anderson impurity models with any number of baths. It captures initial correlations of equilibrated states, and it is controlled in the sense that there is a small parameter that can be tuned in practice in order to approach the exact solution.

Our method is a generalization of the inchworm\cite{Cohen15} quantum Monte Carlo (QMC) method, which was originally formulated on the forward--backward Keldysh contour, to the full Keldysh contour with an additional temperature or imaginary-time branch (also known as the Konstantinov--Perel' contour \cite{Konstantinov1961}). This enables the treatment of interacting equilibrium initial conditions. We also develop a way to obtain nonequilibrium Green's functions, currents, and spectral functions. The Inchworm Monte Carlo algorithm is based on a reformulation of the nonequilibrium hybridization expansion \cite{Muhlbacher08,Schiro09,Werner09,Antipov2016} in terms of bare and interacting atomic state propagators, which are iteratively generated. Its main advantage as compared to other Monte Carlo methods is that the dynamical sign problem,\cite{Schiro09} which causes an exponential amplification of uncertainties as a function of real time, is mitigated or overcome,\cite{Cohen15} providing access to substantially longer times.

\section{Model and Method}
\subsection{Model}
We study the Anderson impurity model with Hamiltonian $H$:
\begin{subequations}
\begin{align}\label{eq:hamilt}
H = & H_D + \sum_{\alpha} H_{\alpha} + H_T,\\
H_D = & \sum_{\sigma}\varepsilon_{d} N_{\sigma} + UN_{\uparrow}N_{\downarrow}, \\ 
H_\alpha = & \sum_{k \sigma}\left(\varepsilon_{k} + \frac{\alpha V(t)}{2}\right) n_{\alpha k\sigma}, \\
H_T = & \sum_{\alpha k\sigma}(\mathcal{V}_k^\alpha c_{\alpha k\sigma}^{\dagger}d_{\sigma}+\mathcal{V}_k^{\alpha *} d_{\sigma}^{\dagger}c_{\alpha k\sigma}).
\end{align}
\end{subequations}
$H$ describes an interacting dot ($H_D$) coupled to two non-interacting leads ($H_\alpha$) by tunneling processes ($H_T$). The dot Hamiltonian $H_D$ spans a Hilbert space generated by $d_\uparrow^\dagger$ and $d_\downarrow^\dagger$ with four `atomic states' $|\phi\rangle$ $=$ $|0\rangle$, $|\uparrow\rangle$, $|\downarrow\rangle$, and $|\uparrow \downarrow\rangle$ and dot occupation $N_\sigma=d^\dagger_\sigma d_\sigma$. $\varepsilon_d$ is the impurity level spacing, and $U$ is the electronic repulsion strength. Lead electrons are characterized by a spin index $\sigma=\uparrow\downarrow$, a momentum quantum number $k$, and a lead index $\alpha =\pm1$, and are annihilated by the operator  $c_{\alpha k \sigma}$. Lead densities are $n_{\alpha k\sigma}=c_{\alpha k\sigma}^\dagger c_{\alpha k\sigma}$ and $\varepsilon_k$ is the lead dispersion.
$\alpha = \pm 1$ labels the left $(+)$ and right $(-)$ leads.  $\mathcal{V}_k^\alpha$ is the tunneling matrix element describing hopping processes between the impurity and the leads. 

We consider two cases: the equilibrium case, where none of the parameters are time-dependent and $V(t)=0$; and the case of a symmetric voltage quench $V(t)=V\theta(t)$, with $\theta(t)$ being a Heaviside step function. In the second case, the system is in equilibrium for $t<0$, and for $t>0$ the lead levels $\epsilon_k$ are instantaneously moved to $\epsilon_k \pm \frac{V}{2}$, with the sign depending on the lead index $\alpha$. We are interested in computing equilibrium and nonequilibrium Green's functions, spectral functions, time-dependent and steady state currents and occupations.

The parameters $\mathcal{V}_k^\alpha$ and $\epsilon_k$ are chosen such that
\begin{align}
\Gamma^{\alpha} \left(\omega\right) = \pi \sum_k |\mathcal{V}_k^\alpha|^2 \delta(\omega - \epsilon_k)
\end{align}
describes a flat band centered at zero with a fermi function like cutoff,
\begin{align}\label{eq:gamma}
\Gamma^{\alpha} \left(\omega\right) = \frac{\Gamma^\alpha}{\left(1+ e^{\nu \left( \omega - D \right)} \right) \left(1 +e^{- \nu \left( \omega + D \right) } \right)}.
\end{align}
Throughout the paper we use $\Gamma^\alpha = \Gamma = 1$; $D = 5$; $\nu = 3$ (unless specified as $\nu=10$); $U=4$ and $U=10$; and temperature $T = 1$.

\subsection{Inchworm Expansion on the Keldysh Contour}\label{sec:inchworm}
We express time-dependent expectation values of operators
\begin{align}
\langle A(t)\rangle=\frac{1}{Z}\text{Tr} \left\{e^{-iHt}e^{-\beta H}e^{iHt}A \right\}, \label{eq:evol}
\end{align}
with $H$ given by Eq.~\ref{eq:hamilt} using a variant of the hybridization expansion quantum Monte Carlo method formulated on the Keldysh contour with forward, backward, and equilibrium (or temperature) branches. The method is based on an expansion of the Hamiltonian into a perturbation series in terms of $H_T$ 
%The hybridization expansion \cite{Werner06} algorithm is a continuous-time algorithm \cite{Rubtsov05,Gull11_RMP} based on expanding the expectation value of operators in powers of the hybridization $\left(\mathcal{V} c^\dagger d + \mathcal{V}^{*} d^\dagger c\right)$.
%
and was originally developed for equilibrium problems, where only the equilibrium branch appears.

As illustrated in Ref.~\onlinecite{Gull10_bold}, the hybridization expansion can be formulated in terms of atomic state propagators $p_\phi$, where $p_\phi (t_1, t_2)$ contains all possible combinations of hybridization events between the two times $t_1$ and $t_2$ that leave the system in atomic state $|\phi\rangle$ at times $t_1$ and $t_2$. Propagators are `contour-causal' by construction, {\it i.e.} all contributions to a propagator $p_\phi (t_1, t_2)$ are given in terms of propagators and hybridization events with time indices between $t_1$ and $t_2$.

Using the cyclic property of the trace we choose a time-ordering on the contour such that the `minus' branch describing the evolution $e^{-iHt_\text{max}}$ from $t_\text{max}$ to $0$ occurs before the equilibrium Matsubara branch from 0 to $-i\beta$, which occurs before the `plus' branch $e^{iHt_\text{max}}$ from 0 to $t_\text{max}$, rather than the usual ordering $e^{-\beta H}e^{iHt}Ae^{-iHt}$, allowing us to insert single-time operators at the beginning or end (rather than in the middle) of the contour.

\begin{figure}[]
\includegraphics[width=\columnwidth,trim={0 5.8cm 0 0},clip]{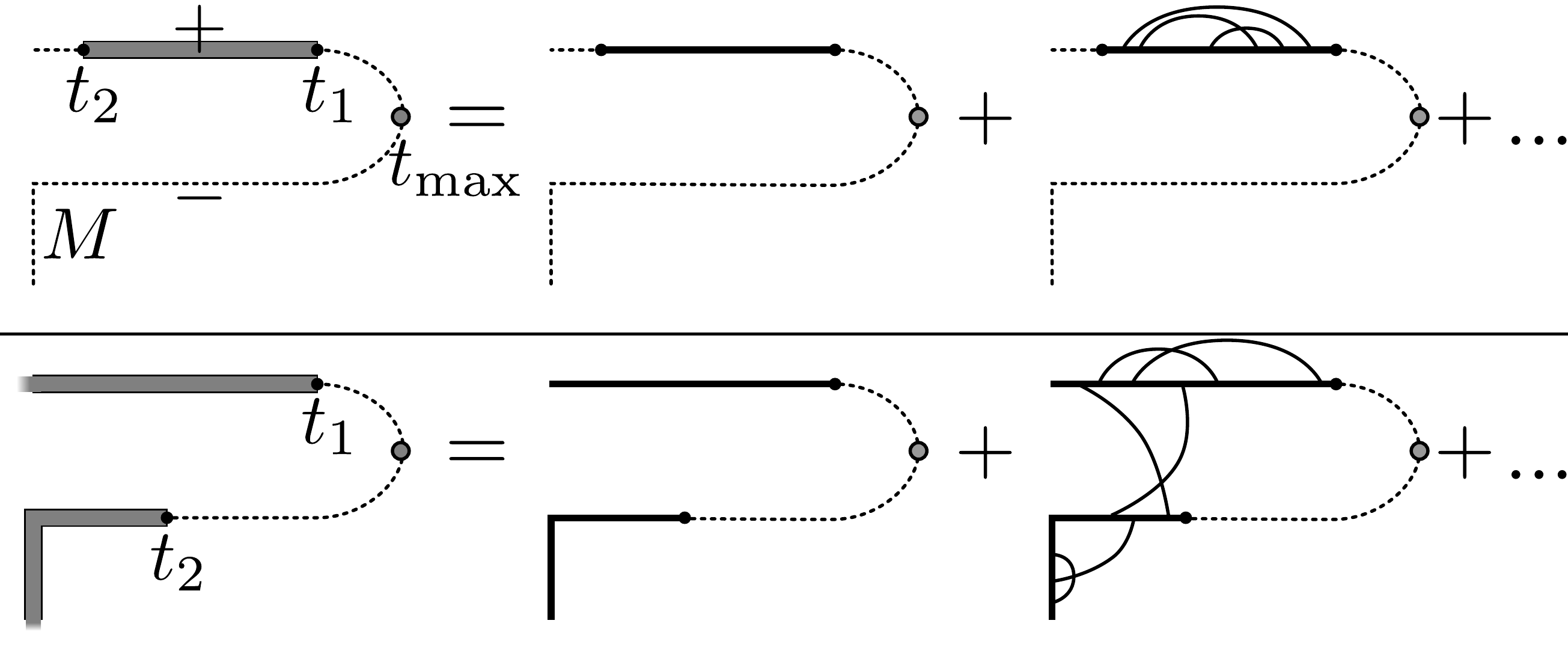}
\caption{Illustration of hybridization expansion diagrams on the Keldysh contour with equilibrium branch. Top panel: Full propagator on the  $+$ branch from $t_2$ to $t_1$ (left diagram) is given by the the bare propagator (middle diagram) plus all possible combinations of hybridization events between $t_2$ and $t_1$, one of which is drawn as the right diagram. Bottom panel: Full propagator spanning the $-$, equilibrium, and $+$ contour containing diagrams that span the contour.
}
\label{fig:bareprops}
\end{figure}
Propagators on the Keldysh contour are illustrated graphically in Fig.~\ref{fig:bareprops} (for formulas see the appendix):  The top panel shows a propagator $p_\phi$ on the upper `+' branch, with $t_1 > t_2$, which is given by the ``bare'' atomic state propagation in the absence of hybridization events (middle diagram) and all possible connected and disconnected combinations of hybridization lines, one of which is shown as the diagram on the right. The bottom panel shows a propagator starting on the lower ``$-$'' branch, propagating along the equilibrium branch, and continuing on the upper ``$+$'' branch. It contains hybridization lines connecting the real time ``$+$'' and ``$-$'' branches to the equilibrium branch, thereby introducing thermal entanglement into the system.

\begin{figure}[]
\includegraphics[width=\columnwidth]{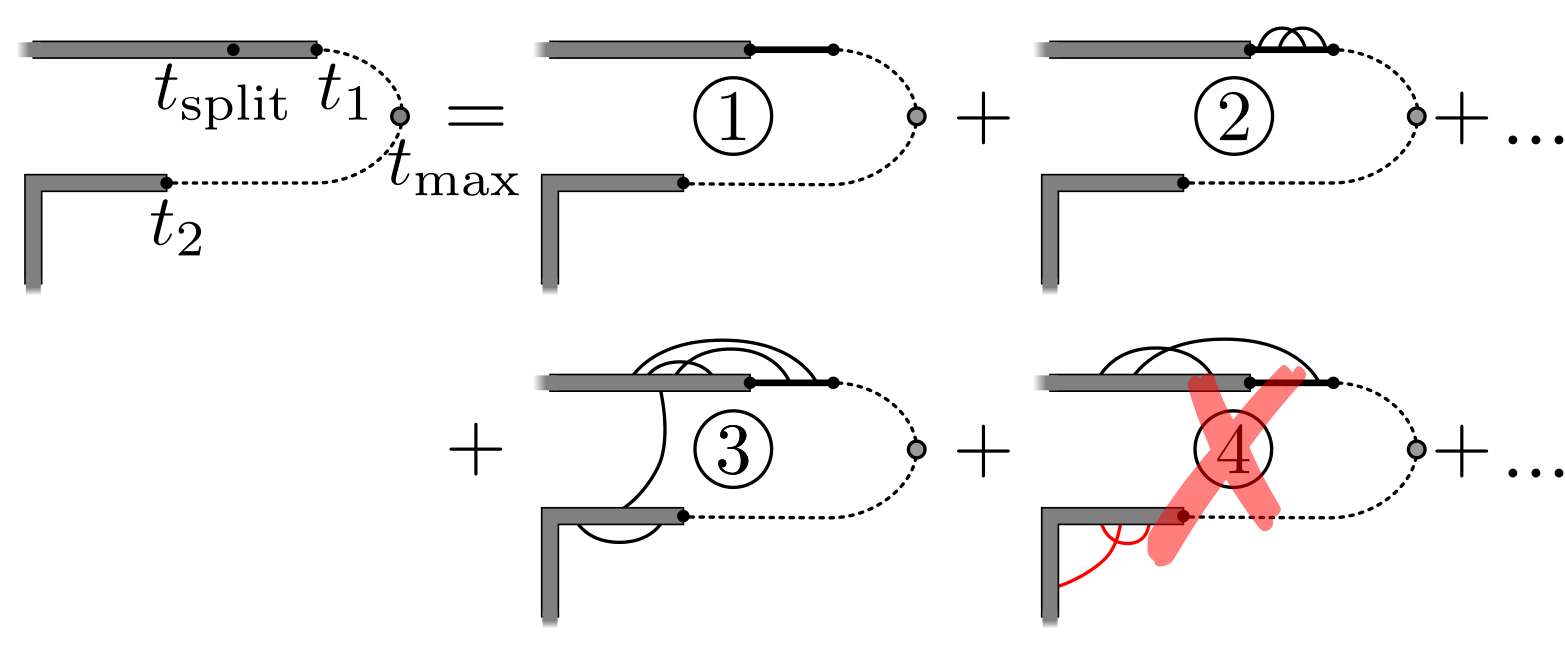}
\caption{Illustration of the diagrams generating a propagator in the inchworm formalism. The propagator from $t_2$ to $t_1$ (left diagram) is, at lowest inchworm order, given by the propagator from $t_2$ to $t_\text{split}$ joined with a bare propagator from $t_\text{split}$ to $t_1$ \textcircled{1}. In higher order diagrams, hybridization lines are either contained in the region $t_\text{split}$ and $t_1$ as in \textcircled{2}; or, connected by crossing to an endpoint in that region, as in \textcircled{3}. Diagrams with inclusions not obeying this rule, such as \textcircled{4}, are already included in the inchworm propagator and should not be summed over.
}
\label{fig:inchprops}
\end{figure}
The contour-causality of the propagators can be used to construct propagators over longer time intervals from previously computed propagators over shorter intervals. The concept is illustrated in Fig.~\ref{fig:inchprops}, which shows how a propagator from $t_2$ to $t_1$ can be expressed as a propagator from time $t_2$ to $t_\text{split}$ multiplied by the bare propagation from time $t_\text{split}$ to $t_1$ (diagram \textcircled{1}), supplemented by diagrams which have hybridization events between time $t_\text{split}$ and $t_1$. Of those, diagram \textcircled{2} only has hybridization events between $t_\text{split}$ and $t_1$; diagram \textcircled{3} has hybridization events starting between $t_\text{split}$ and $t_1$ and reaching backward in time to a position between $t_2$ and $t_\text{split}$, along with additional hybridization lines that cross those lines. In contrast, diagram \textcircled{4} contains a separate cluster of hybridization lines (red online) which are already contained in a propagator, and is therefore not part of the series of diagrams to be summed in order to construct the propagator over the full interval.

Computing corrections to known propagators rather than computing the entire propagator at once is efficient if the propagator consists of many short clusters of hybridization lines, so that most of the interaction contribution can be absorbed in previously computed propagators. In the case of the forward--backward contour, it was shown that at least for some parameters this procedure changed the scaling from exponential to polynomial, overcoming the dynamical sign problem.\cite{Cohen15}

\begin{figure*}[t]
\includegraphics[height=5.5cm]{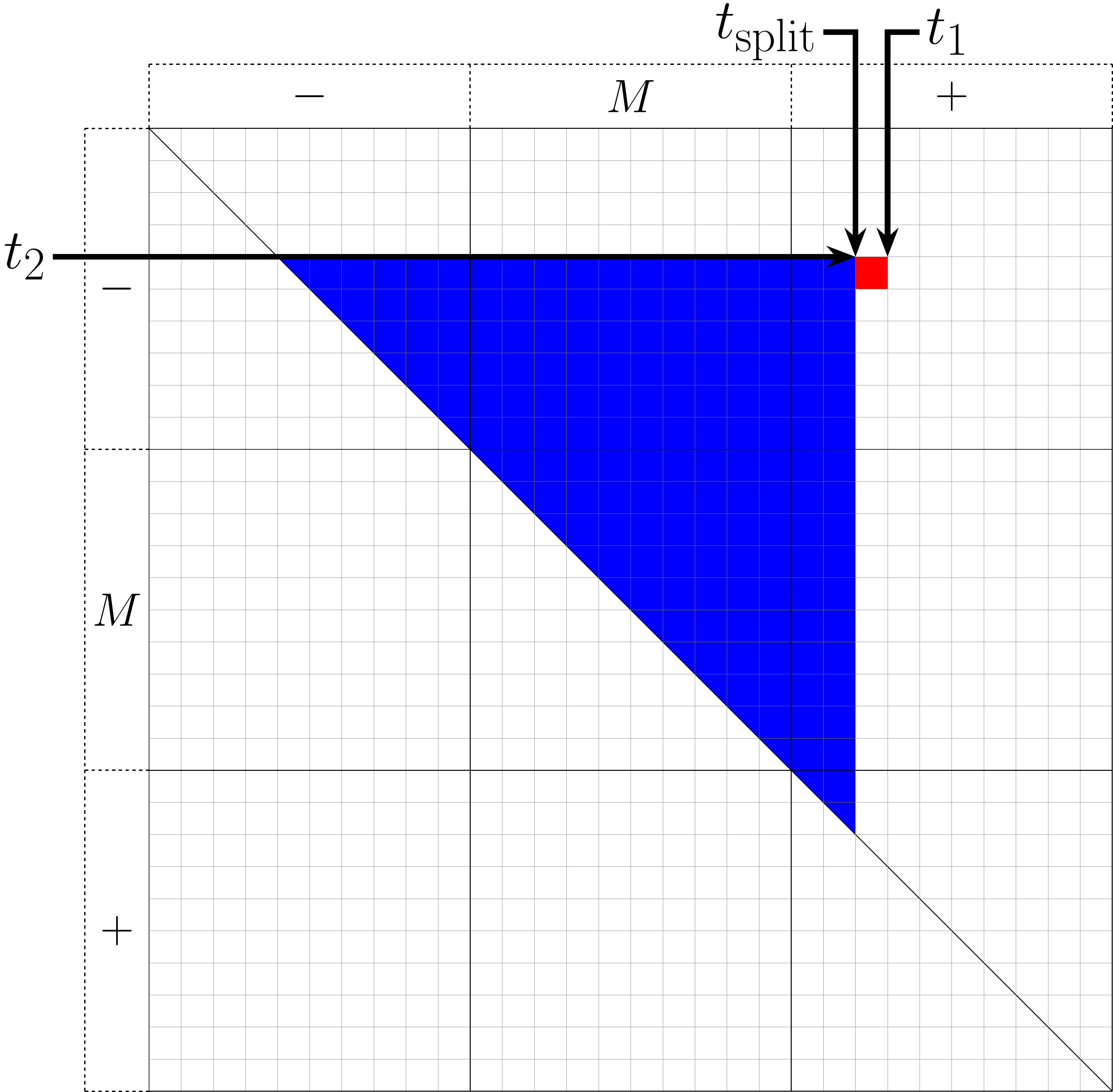}
\hfill
\includegraphics[height=5.5cm]{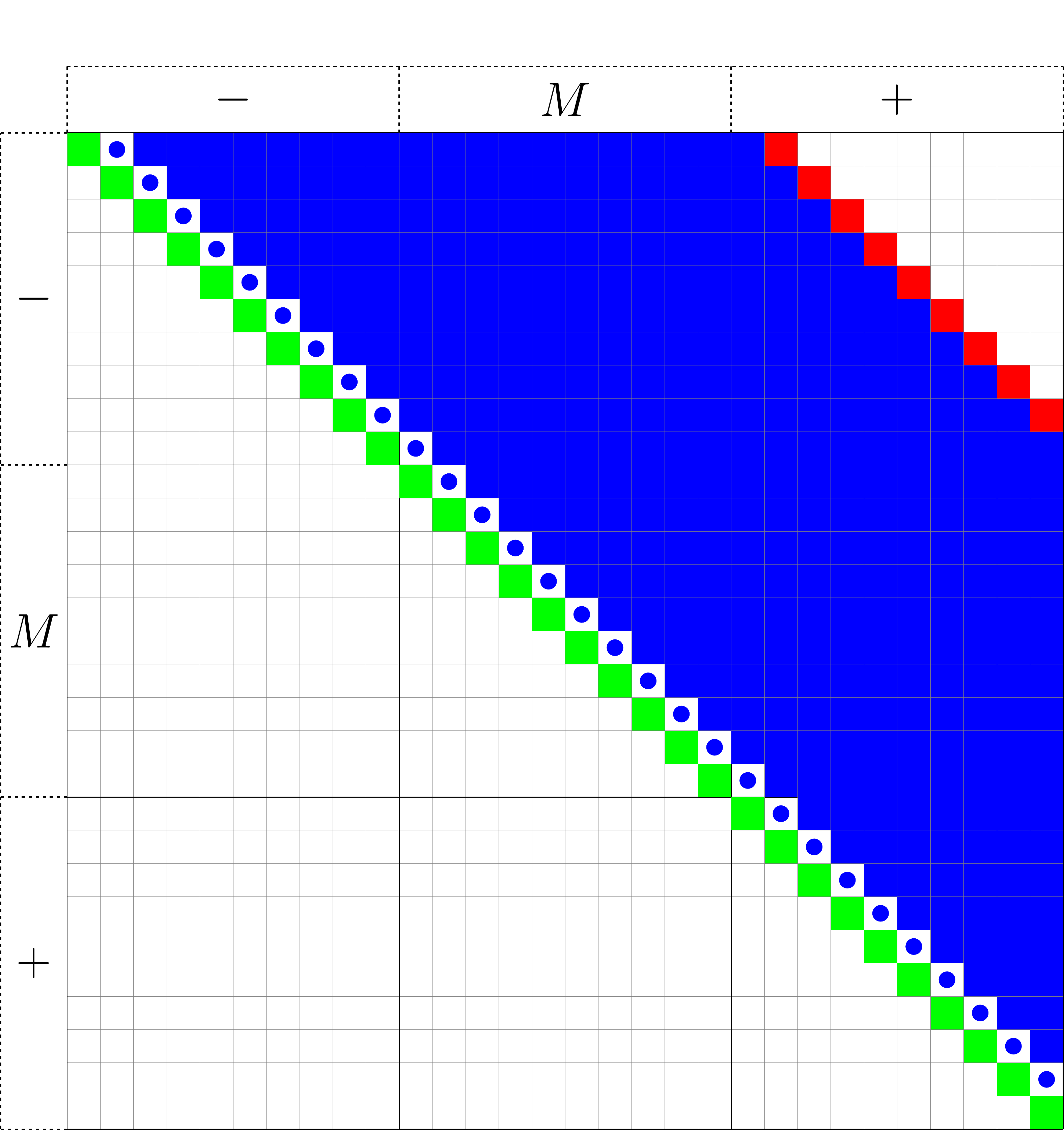}
\hfill
\includegraphics[height=5.5cm]{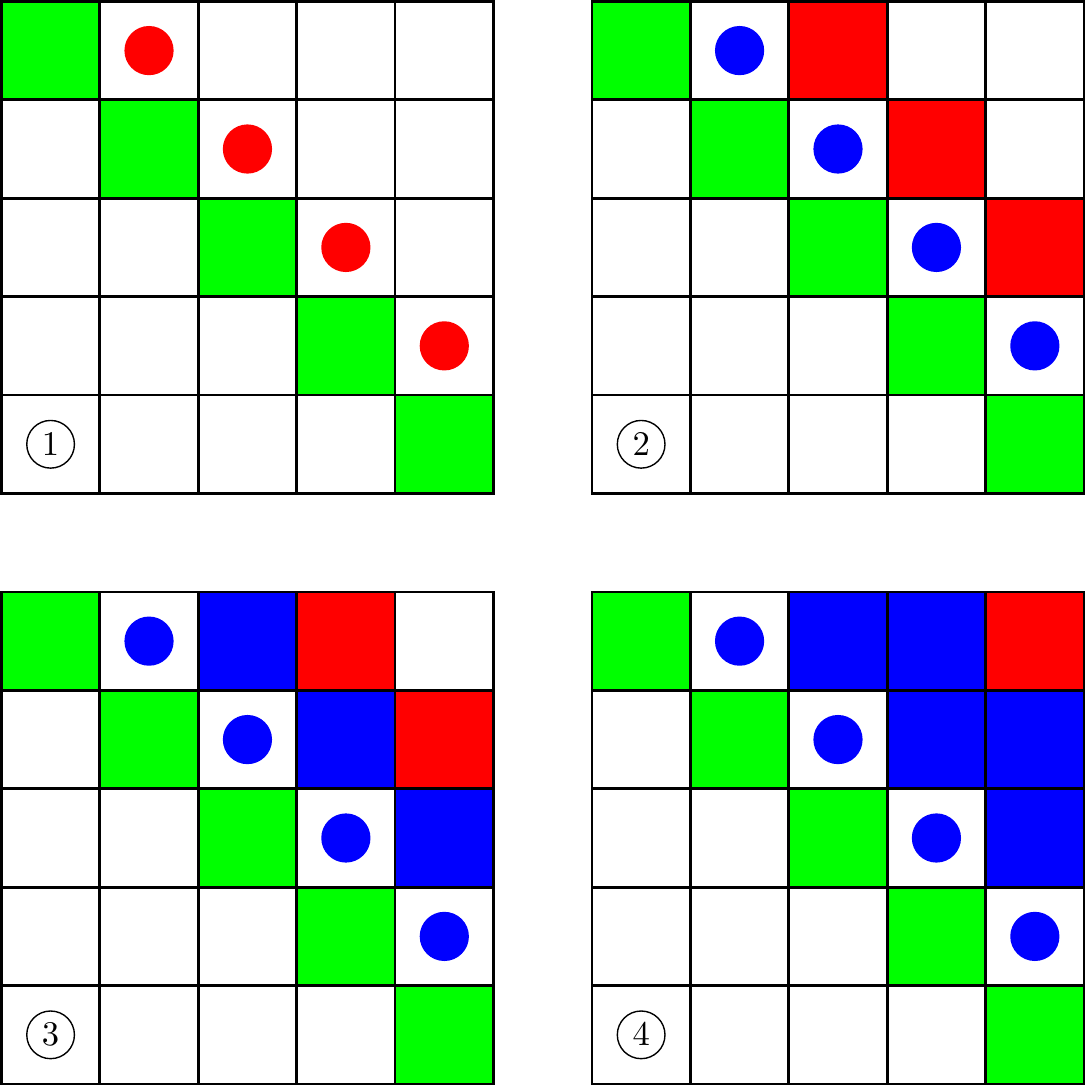}
\caption{
Left panel: Illustration of the causal structure of the inchworm propagators. A propagator $p_\phi(t_1, t_2)$ (red) depends on all propagators with start point on or after $t_2$ and end point up to $t_\text{split}$ (blue).
Middle panel: Illustration of the inchworm algorithm showing the values known at some step (blue) and values that can be computed using known values in blue (red). Evaluation of the full propagator proceeds diagonally towards the upper right corner from the initial values as dictated by the causal structure of the propagators.
Right panel: step-by-step illustration of the algorithm. Cells with circles are evaluated with the bare algorithm and solid cells with inchworm. Red cells represent propagators computed at a given step, while blue cells represent already computed propagators. Green cells in both panels are trivial and can be evaluated analytically.
}
\label{fig:task_matrix}
\end{figure*}

As evident in Fig.~\ref{fig:inchprops}, computing the propagator from time $t_2$ to time $t_1$ requires knowledge of all propagators from $t''$ to $t'$, with $t_2 < t'' < t' < t_1$. 
The left panel of Fig.~\ref{fig:task_matrix} shows the set of known propagators (blue) needed to compute a new propagator (red), in this case with start time on the ``$-$'' contour and end time on the ``$+$'' contour. This step, which we call ``inching'', could be repeated any number of times to generate longer propagators from sets of shorter ones. This suggests an iterative algorithm, graphically illustrated in the middle panel of Fig.~\ref{fig:task_matrix}: the inchworm quantum Monte Carlo method.\cite{Cohen15}

The algorithm begins with the discretization of the backward, equilibrium, and forward branches of the Keldysh contour into $N$ equidistant time slices with distance $\Delta t$, enumerated by $k=1$, $\dots$, $N$.
Same-time propagators between identical contour times (which lie on the main diagonals of the matrices in Fig.~\ref{fig:task_matrix}) are trivial, and can be evaluated analytically. In the right two panels of Fig.~\ref{fig:task_matrix}, these propagators are shown in green. Next, propagators for short time differences $n\Delta t$ (with $n$ a small integer chosen to be large enough that smoothly interpolated functions can be obtained) can be computed straightforwardly using the bare QMC method, which is efficient for short times. There are $Nn$ such propagators, with time arguments between times $k\Delta t$ and $(k+n) \Delta t$. In the right panel of Fig.~\ref{fig:task_matrix}, the minimal number $n=1$ is taken and the result is shown as cells with circles.
In the next step and each following step, propagators from any time $k\Delta t$ to any time $(k + n + 1)\Delta t$ need to be computed. We obtain them by stochastically generating all diagrams illustrated in Fig.~\ref{fig:inchprops}, where $t_2$ is $k\Delta t$, $t_1$ is $(k+n+1)\Delta t$, and $t_\text{split}$ is set to $(k+n)\Delta t$.
We then increase $n$ by 1 and iterate the last step of the procedure, simulating propagators from $k\Delta t$ to $(k+n+1)\Delta t$ based on any propagator with times between $k\Delta t$ and $k+a \Delta t$ at each iteration, until the propagator from time $0$ to time $N\Delta t$ is generated (top right corner of Fig.~\ref{fig:task_matrix}).

The right panel of Fig.~\ref{fig:task_matrix} shows the inchworm procedure step by step, with the elements participating in each particular step highlighted in red. This illustrates the trivial elements (in green), the initial bare step (circles), and two inchworm steps (solid) that gradually extend the known propagators along the Keldysh contour.

The complete procedure requires performing $O(N^2)$ interdependent Monte Carlo simulations of diagrams, each of which is represented by a cell in the matrices in Fig.~\ref{fig:task_matrix}. However, at any given step all the computations corresponding to cells which can currently be evaluated---\textit{e.g.}, the cells colored in red at every step of the right panel of Fig.~\ref{fig:task_matrix})---can be evaluated simultaneously and independently. Furthermore, since the computation of each individual cell is a \textit{regular} QMC simulation, the work it entails can also be trivially split between any number of compute nodes. The inchworm algorithm therefore lends itself to extremely efficient parallelization strategies. However, since after every step at least some data synchronization between cells is required, it is not `embarrassingly parallel' in the sense of standard Monte Carlo methods.

Each individual inchworm step is exact for any $\Delta t$, provided that all intermediate propagators are exactly known. In practice the propagators, generated by previous inchworm steps or a bare calculation, are interpolated on a grid with discretization $\Delta t$. This discretization introduces errors for large $\Delta t$, especially where propagators change on a time scale comparable to $\Delta t$, and needs to be controlled by extrapolating to $\Delta t\rightarrow 0$.

\begin{figure}[]
\includegraphics[width=\columnwidth]{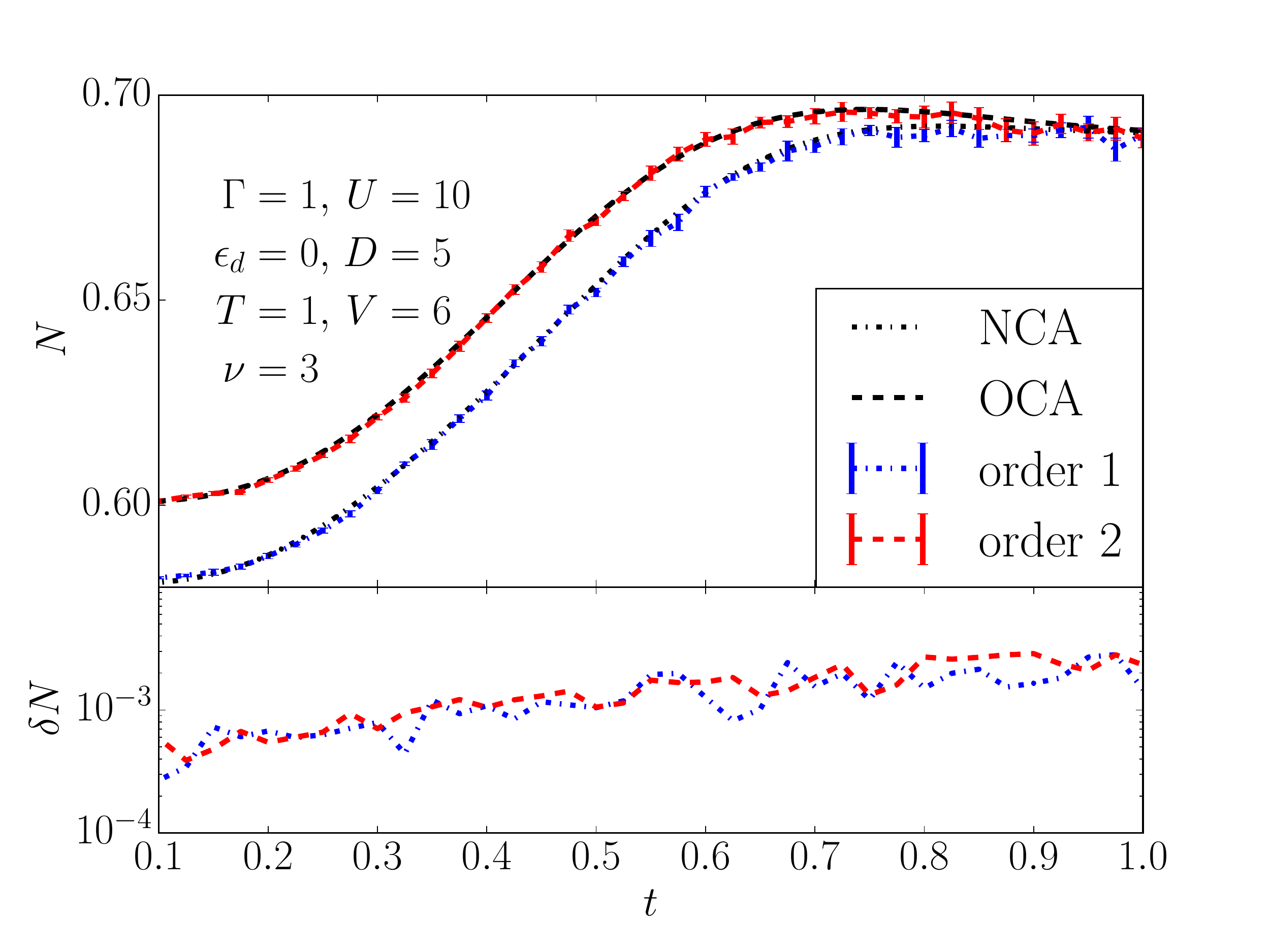}
\caption{
Top panel: Time evolution of the impurity occupation $N$ after a voltage quench using the non-crossing and one-crossing approximations (NCA and OCA, respectively). Black lines: semi-analytically computed NCA and OCA solutions. Blue line: NCA solution generated from an inchworm expansion truncated to order one. Red line: OCA solution from an inchworm expansion truncated to order 2. 
Bottom panel: Statistical error estimate of the quantities shown in the upper panel.
}
\label{fig:ordercomp}
\end{figure}

In the limit $\Delta t \rightarrow 0$, at most one hybridization event will occur between $t_\text{split}$ and $t_1$. In this limit, the method simplifies to the well-known semi-analytic `$N$-crossing' approximations when inchworm diagrams are restricted to low orders.\cite{Cohen15} At truncation to order $n=1$, NCA results\cite{Keiter70,Bickers87} are obtained. Truncation to order $n=2$ yields the OCA,\cite{Pruschke89} the two-crossing approximation (2CA) is generated for order $n=3$, etc. Fig.~\ref{fig:ordercomp} illustrates agreement within error bars of numerical results for the propagators truncated to $n=1$ and $n=2$ to the NCA and OCA approximations. Below, it is shown that the size of the inchworm error does not strongly depend on time. This implies that for `crossing' expansions on the order of the OCA and above, inchworm Monte Carlo provides an efficient alternative to the direct integration of the equations of motion.

Continuous-time QMC requires the sampling of diagrams to all orders. In bare expansions very high order diagrams are easily sampled, because (due to Wick's theorem) the sum over all diagrams for a particular configuration of order $2k$, of which there are $k!$, can be written as the determinant of a $k\times k$ matrix and evaluated at polynomial scaling using linear algebra algorithms.\cite{Gull11_RMP} However, in bold and inchworm Monte Carlo a factorial number of diagrams must be explicitly summed over at each order, and the cost of enumerating these diagrams quickly becomes prohibitive (evaluating the sum over permutations stochastically leads to a sizable increase in the overall sign problem). We therefore truncate the series at a predetermined maximum order and observe convergence as that order is increased. This corresponds to observing convergence in the hierarchy NCA $\rightarrow$ OCA $\rightarrow$ 2CA $\cdots$, each of which contains an infinite subseries of all the bare diagrams which extends to infinite order. In this work, we typically truncate this hierarchy at order 5--7.

\subsection{Normalization and Wang--Landau}\label{sec:wl}
Additional technical complications arise when the inchworm algorithm is extended to the full Keldysh contour with the imaginary time branch. Monte Carlo importance sampling does not generate absolute values of observables. Rather, it generates probability ratios, or values up to an unknown normalization constant. This normalization can be computed by comparing to a known reference, e.g. a zero or first order diagram, as long as the overlap of the series with that reference is large. In systems where the low-order diagrams are not important, the overlap with the reference becomes small, causing a variance problem. This problem can be solved by changing the sampling such that regions at low order are visited more often using a generalized ensemble technique. We chose to modify our sampling using the Wang--Landau algorithm\cite{Wang01,Wang02} to generate a flat histogram in expansion orders. These algorithms, originally designed to overcome ergodicity barriers at first order phase transitions, were previously extended to quantum phase transitions\cite{Troyer03} and applied to CT-QMC\cite{Li09} to compute thermodynamic potentials and overcome ergodicity problems.\cite{Gull11_RMP,Iskakov16}

\begin{figure}[]
\includegraphics[width=\columnwidth]{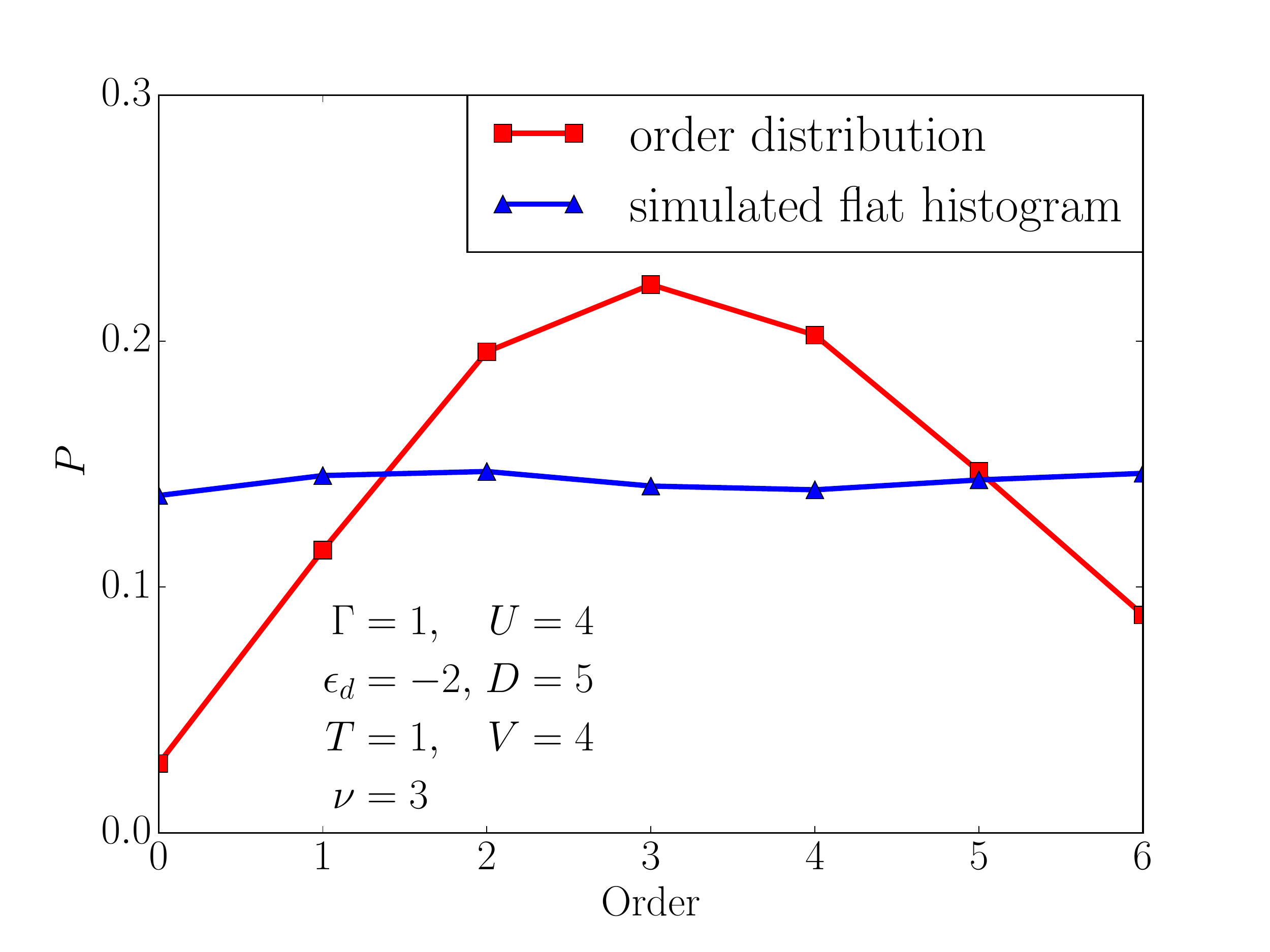}
\caption{Order distribution (red) and sampled ``flat'' histogram (blue) for a Wang--Landau simulation of the inchworm propagators. The large overlap of the reweighted distribution with order zero allows normalization to low order diagrams.}\label{fig:WangLandau}
\end{figure}
Fig.~\ref{fig:WangLandau} shows the expansion order histogram of an inchworm propagator simulation for a representative set of parameters $\Gamma=1$, $U=4$, $\epsilon_d=-2$, $D=5$, $T=1$ and $V=4$ up to order 6. The red line shows the contribution of the absolute value of the diagrams at each order to the inchworm propagator. It is evident that diagrams at higher order acquire higher weight, and that diagrams near order zero are strongly suppressed, making normalization to low order diagrams difficult. If, in contrast, the sampling weights are changed to produce a `flat' order distribution (see blue line in Fig.~\ref{fig:WangLandau}), each expansion order is visited equally often, and normalization to low order diagrams is possible. 

\begin{figure}[]
\includegraphics[width=\columnwidth,trim={0 0 0 11.4cm},clip]{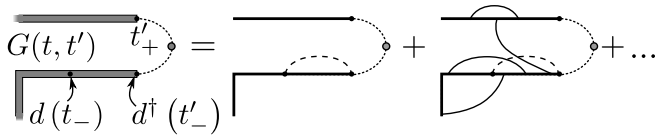}
\caption{Illustration of the diagrams generating a Green's function. The Green's function $G(t,t')$ is given at lowest order by the middle diagram, where the dashed line is a virtual hybridization line from the Green's function creation/annihilation operators. Higher order diagrams contain at least one hybridization line that crosses the virtual line. A sample term is shown in the right diagram.}\label{fig:inchgf}
\end{figure}

\subsection{Observables: single-time observables, Green's functions, and currents}
We are predominantly interested in obtaining observable estimates for  densities, currents, and Green's functions. The formalism introduced in Ref.~\onlinecite{Cohen15} provides a direct way to obtain diagonal elements $\rho_{jj}$ of the local density matrix $\rho$, such as magnetizations and (spin-) densities by evaluating propagators at equal times on `plus' and `minus' branch of the contour (see appendix for detailed equations).

Currents and Green's functions are two-time observables that cannot be obtained from knowledge of just the population propagators.
A current or Green's function diagram is illustrated in Fig.~\ref{fig:inchgf}. As shown in Ref.~\onlinecite{Werner10}, the expansion for the current and Green's function is given by all possible contractions of hybridization lines in the presence of two additional local operators. 
In terms of the interacting propagators obtained in Fig.~\ref{fig:bareprops}, the Green's function is given, to lowest order, by a product of propagators (middle diagram of Fig.~\ref{fig:inchgf}). 
Higher order corrections consist of hybridization lines crossing the two Green's function operators, and all possible additional crossing lines. 
The right panel of Fig.~\ref{fig:inchgf} shows an example of one such diagram. In contrast to the case of single-time propagators, where we iteratively construct diagrams at longer times using diagrams at shorter times, here we generate all Green's function diagrams at once, simulating them in parallel. 
This procedure works well at high temperatures, where expansion orders stay small.  As $T$ is lowered, diagrams at higher order contribute and an exponential scaling in $T$ is recovered. 

\begin{figure}[]
\includegraphics[width=\columnwidth]{{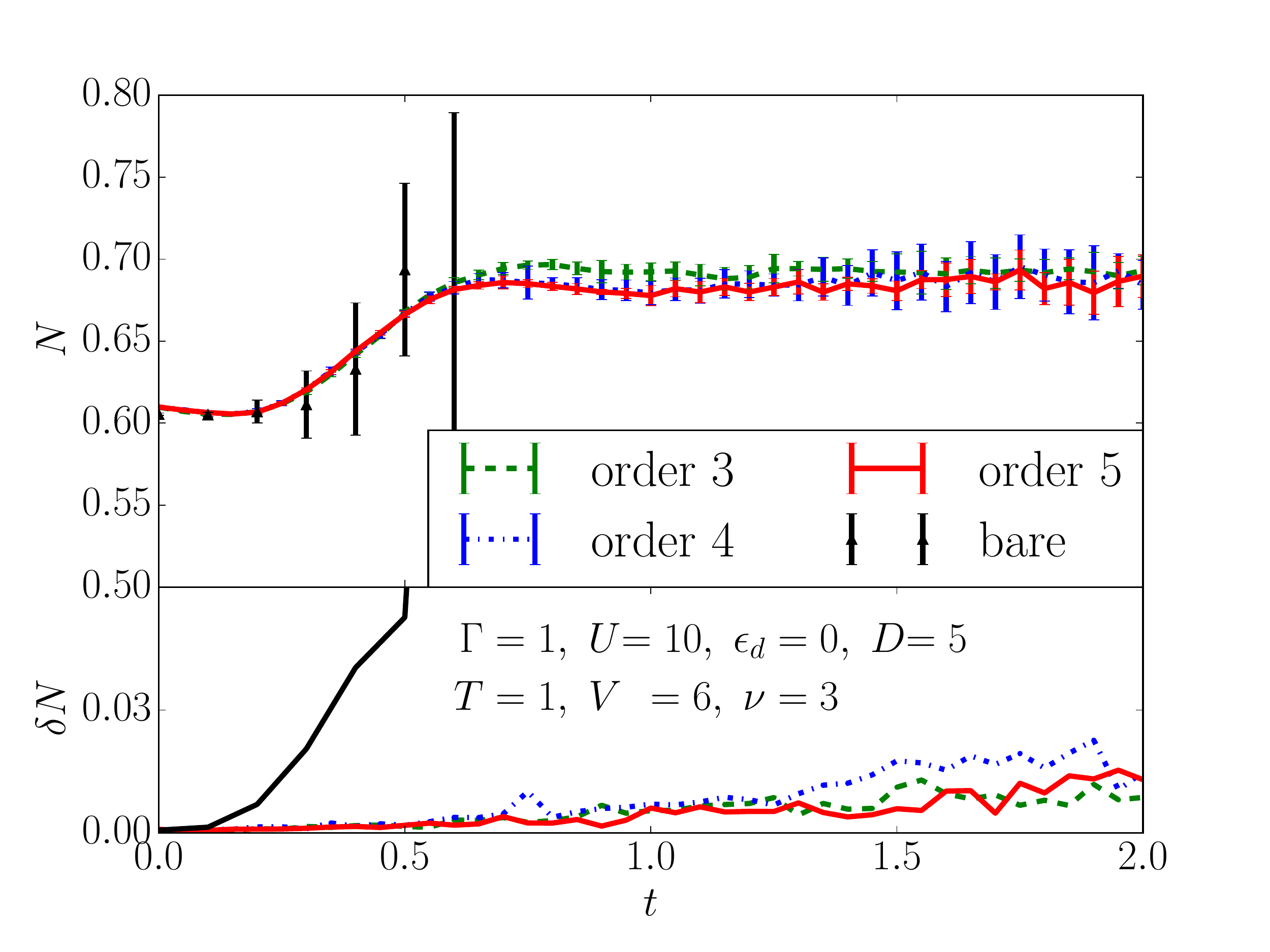}}
\caption{Top panel: Time evolution of the density on the impurity after a voltage quench with $\Gamma= 1$, $U = 10$, $\epsilon_d = 0$, $D = 5$, $T = 1$ and $V = 6$. Results obtained from a bare QMC calculation are shown for $t \leq 0.6$. The inchworm results with different orders agree with the bare result for $t \leq 0.6$ and coincide with each other for longer times. Bottom panel: Error estimates. Data obtained using the bare method shows an exponential increase of the errors as a function of time, whereas inchworm errors grow slower as a function of time.
}\label{fig:population}
\end{figure}

\section{Results}
\subsection{Population and magnetization}
The top panel of Fig.~\ref{fig:population} shows results for the time-evolution of the density after a voltage quench of an impurity with parameters $\Gamma= 1$, $U = 10$, $\epsilon_d = 0$, $D = 5$, $T = 1$ and $V = 6$. Black triangles denote values obtained in a bare QMC simulation, and colored lines the inchworm results with respective maximum order constraints of order $3$, $4$, and $5$ as labeled in the plot.
At short times ($t \leq 0.6$ in these units), the inchworm results agree with the bare calculation within error bars, but for $t \gtrsim 0.3$ the bare QMC error bars are too large to be useful. Inchworm results for orders $4$ and $5$ coincide within error bars at long times, indicating that a solution obtained within a three-crossing approximation calculation would be accurate.
The bottom panel of Fig.~\ref{fig:population} shows statistical error bars for the data shown in the top panel. Errors for the bare calculation increase exponentially as a consequence of the dynamical sign problem.
In contrast, the statistical inchworm error estimate grows slowly, allowing access to significantly longer times. We note that in order to account for error propagation and non-linear cross-correlations from short-time propagators to long-time propagators within the inchworm algorithm, the error bars have been obtained by running multiple (in this case eight) complete independent calculations, each of which includes a different realization of the statistical noise at all times. The standard deviation between the different runs provides a useful estimate of the confidence interval, whereas the standard deviation within each run---which does not account for error propagation---grossly underestimates the error.

It is remarkable that no exponential growth of the errors is seen, signaling that the dynamical sign problem has been overcome. However, a gradual, approximately linear increase of errors with time is visible.

\begin{figure}[]
\includegraphics[width=\columnwidth]{{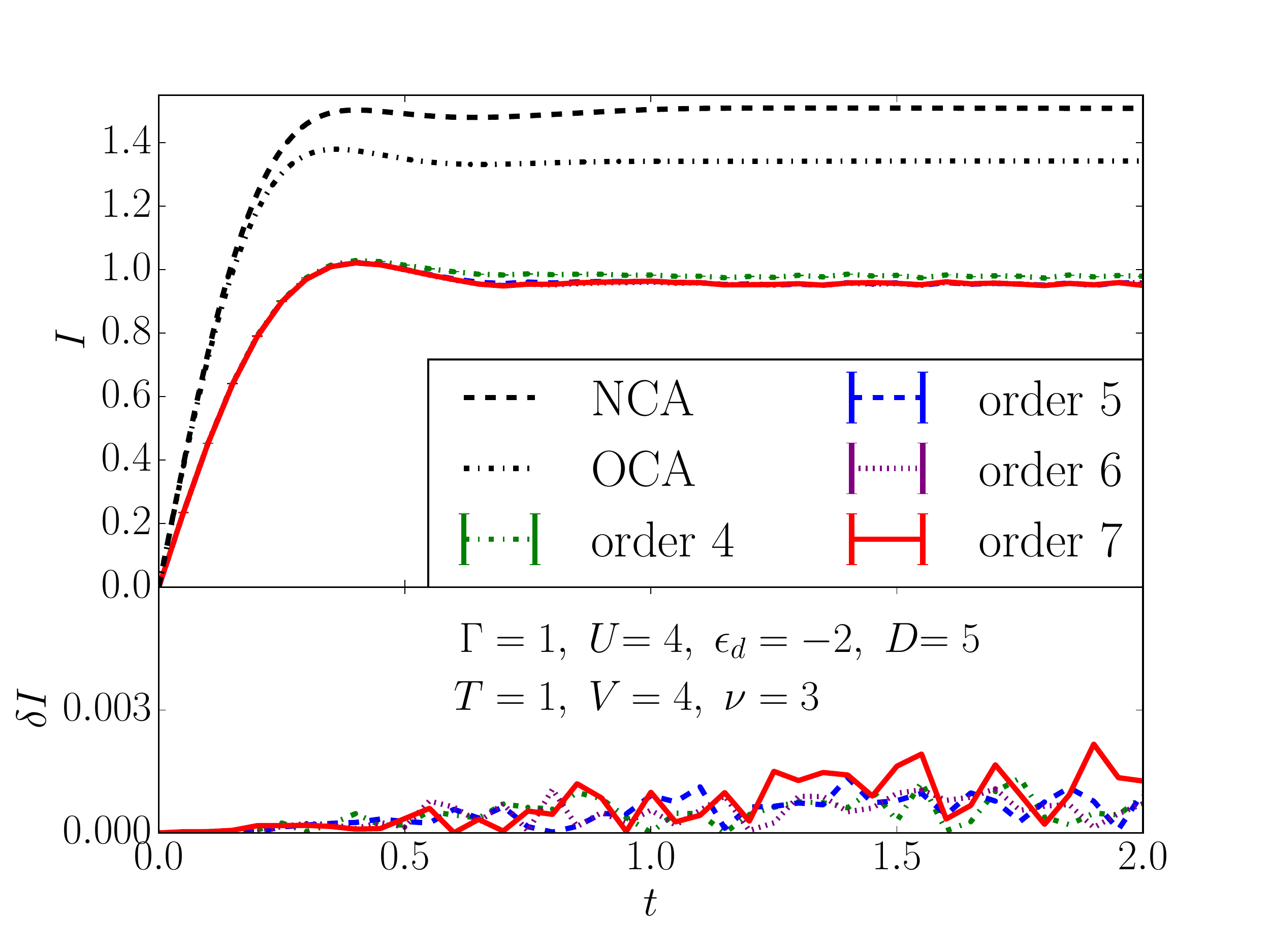}}
\caption{Top panel: The current dynamics after a voltage quench with $\Gamma= 1$, $U = 4$, $\epsilon_d = -2$, $D = 5$, $T = 1$ and $V = 4$. The inchworm results with different orders converge as max-order increases. Bottom panel: Error estimates of inchworm data obtained by averaging eight independent calculations. Errors increase as a function of time but avoid the exponential amplification seen in bare calculations.
}
\label{fig:current}
\end{figure}
\subsection{Current}
Fig.~\ref{fig:current} shows results for the time dependence of a current passing through the impurity after a voltage quench from a thermalized equilibrium state. Parameters are $\Gamma= 1$, $U = 4$, $\epsilon_d =-2$, $D = 5$, $T = 1$ and $V = 4$. In the top panel, we observe that both NCA and OCA produce qualitatively wrong results for both the transient and long-time response. 
In contrast, inchworm results at orders 5--7 are in excellent agreement with each other, and order 4 is within about a percent from the converged result. Convergence at order 5 is well within reach of inchworm calculations but far beyond what could be realistically treated with semi-analytical methods. The bottom panel shows a rough estimate of the statistical error of the data shown in the top panel, obtained from the standard deviation of eight independent simulations of this problem. As observed for the densities, the inchworm error grows sub-exponentially in time and order constraint, indicating that the algorithm is able to overcome the dynamical sign problem.

\begin{figure}[]
\includegraphics[width=\columnwidth]{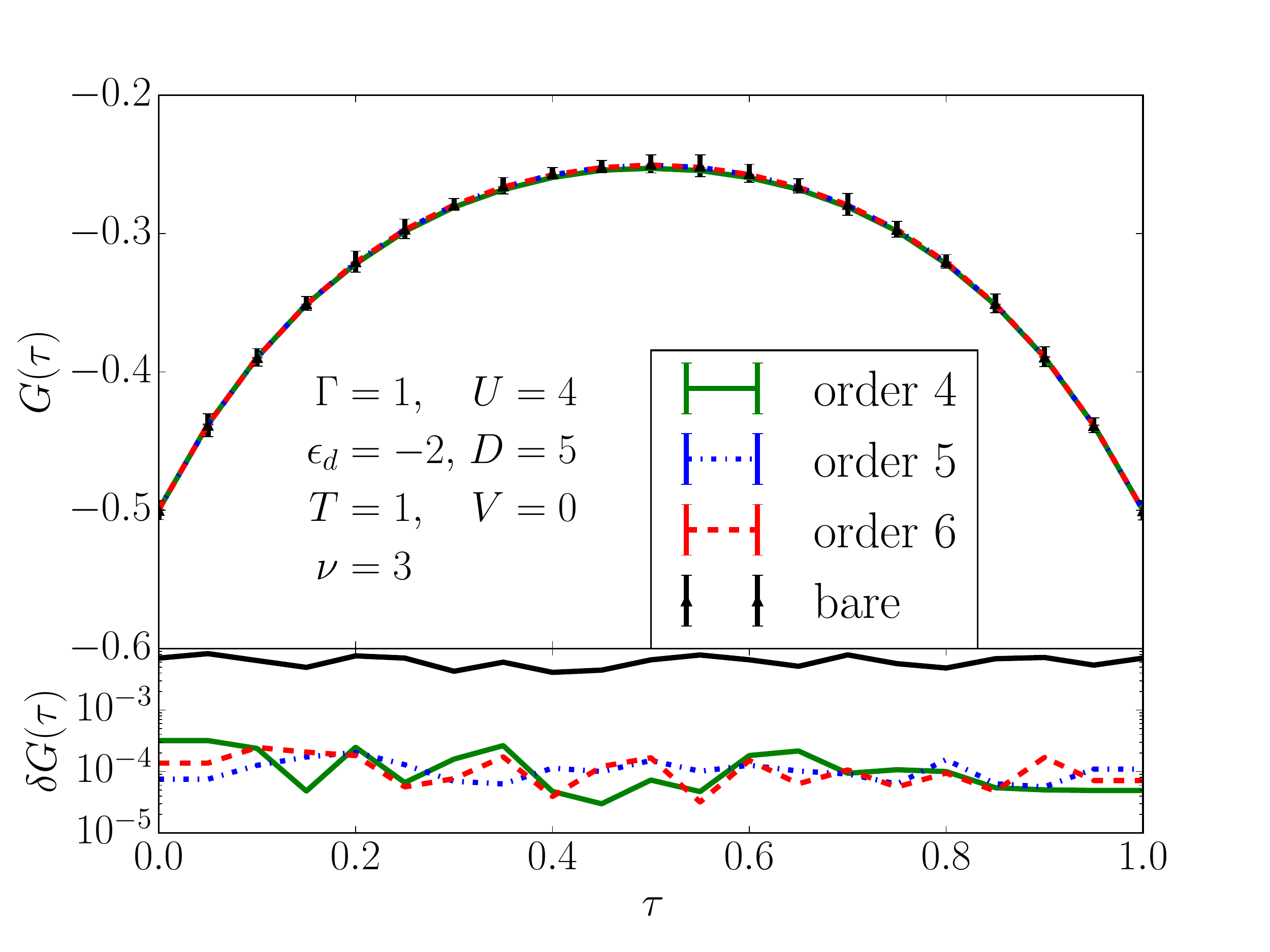}
\caption{Top panel: The imaginary time Green's function in equilibrium (half-filling) with $\Gamma= 1$, $U = 4$, $\epsilon_d = -2$, $D = 5$, $T = 1$ and $V = 0$. Inchworm results with different orders all coincide and agree with the bare calculation. Bottom panel: The error estimate for the inchworms data is approximately constant in imaginary time.
}
\label{fig:gtaueq}
\end{figure}

\subsection{Green's function}\label{sec:gf}
Simulation of diagrams as shown in Fig.~\ref{fig:inchgf} enable both the simulation of currents and of two-time Green's functions. On the full Keldysh contour, a total of nine different types of Green's functions exist. One of them, the imaginary time Green's function, is shown in Fig.~\ref{fig:gtaueq}. The parameters used are $\Gamma= 1$, $U = 4$ and $\epsilon_d = -2$ (such that the system is at half filling), $D = 5$, $T = 1$, and $V = 0$.

As is visible in the upper panel, orders $4$, $5$, and $6$ agree perfectly within error bars with the result obtained by a bare reference hybridization expansion calculation.

Statistical error bars, which do not estimate the systematic errors caused by the order truncation, are shown in the lower panel of Fig.~\ref{fig:gtaueq}. These errors are on the order of $10^{-4}$.

The remaining components of the Green's function are similarly obtained by simulating the diagrams of Fig.~\ref{fig:inchgf}.

\begin{figure}[]
\includegraphics[width=\columnwidth]{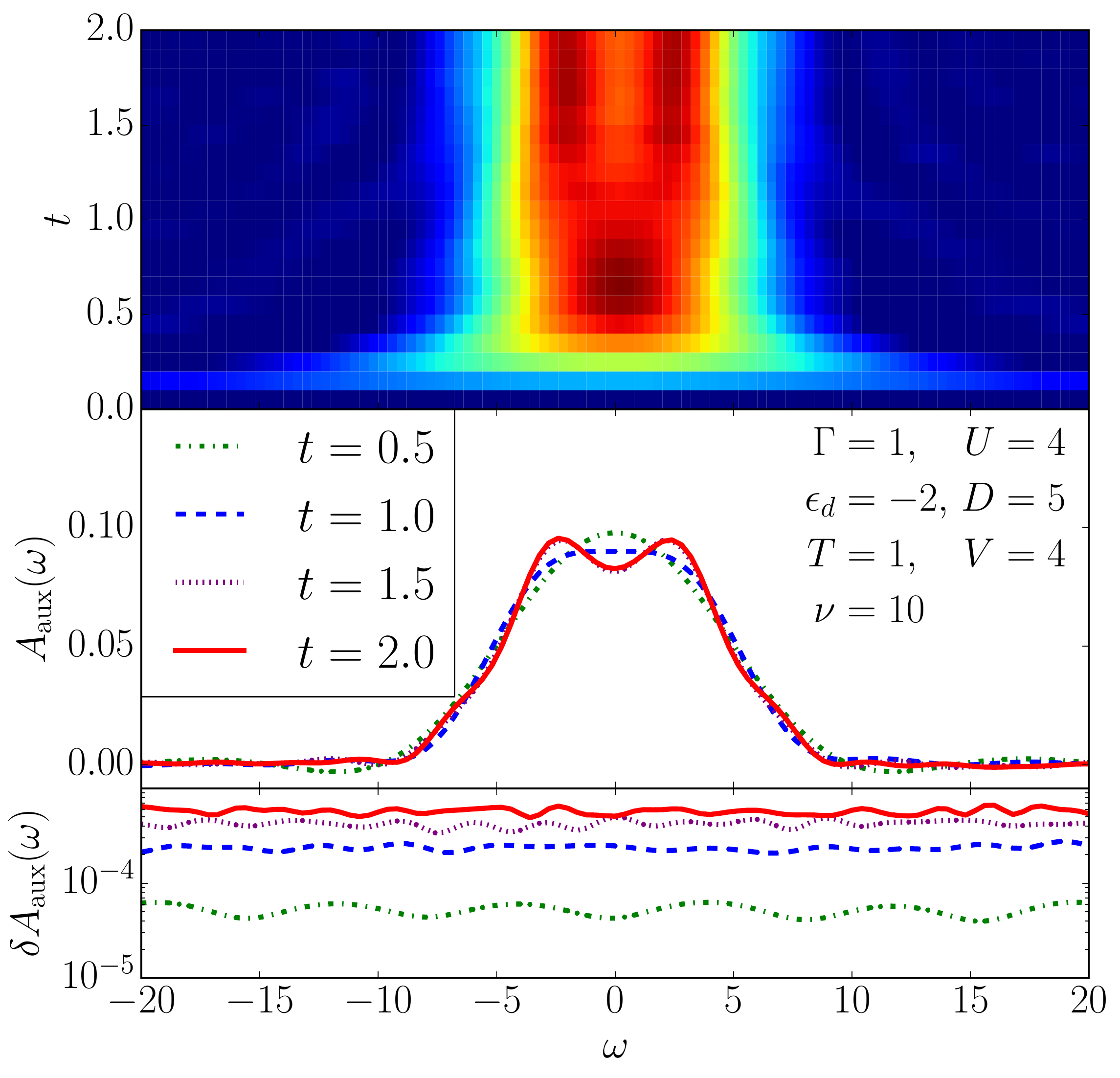}
\caption{Top panel: A contour plot of the dynamics of auxilary current spectrum $A_\text{aux}(\omega, t)$ after a voltage quench with $\Gamma= 1$, $U = 4$, $\epsilon_d = -2$, $D = 5$, $T = 1$ and $V = 4$. The maximum order cut-off for the inchworm calculation is $6$. A formation and a splitting of the Kondo peak are observed as a function of time. Middle panel: Slices of auxiliary current spectrum at different times from the top contour plot. A clear splitting of the spectrum is shown. Bottom panel: Error estimate on the spectral function obtained from eight independent simulations.}\label{fig:auxA}
\end{figure}

\subsection{Steady state spectral function}
Knowledge of Green's functions and currents makes the calculation of interacting single-particle spectral functions possible. Ref.~\onlinecite{cohen_greens_2014-1} introduced a method for computing steady state spectral functions $A(\omega)$ by obtaining steady state currents in two narrow auxiliary leads attached at frequency $\omega$. 
Fig.~\ref{fig:auxA} shows the result in the spirit of the auxiliary lead scheme, but generalized to the full Keldysh contour (see Appendix): initially, at $t=0$, no current is flowing. As the voltage in the main leads, along with the auxiliary lead voltage, is instantaneously switched on, an auxiliary current starts flowing and relaxes on a time scale of about $1.5$ to $2$.

The upper panel shows the time-evolution of this current as a false-color contour plot. The vertical axis is time, the horizontal axis is frequency and the color represents the value of the auxiliary spectral function $A(\omega)$ obtained from the auxiliary currents. This quantity is equivalent to the physical spectral function at the long time limit. The middle panel shows frequency cuts through these data, illustrating a buildup of a more-or-less featureless spectral function at intermediate times ($t=0.5, t=1.0$), which splits into two sub-peaks (associated with the onset of Kondo physics\cite{cohen_greens_2014,anders_steady-state_2008}) as time is extended towards time $t=1.5$ and $2.0$ By time $t=1.5$, all features are converged.

In this parameter regime, both the final steady state spectral function and the time-scale on which results converge are comparable after a quench from an equilibrium thermal state and after a quench from a decoupled initial state,\cite{cohen_greens_2014,cohen_greens_2014-1} illustrating that in this case the presence of equilibrium correlations in the initial state did not substantially accelerate convergence.

The bottom panel shows the statistical errors of these data, obtained by computing the standard deviation of numerical data from several independent calculations. It is clearly visible that as $t$ is increased, errors increase. However, the errors do not increase exponentially, again hinting that the dynamical sign problem has been overcome.

\begin{figure}[]
\includegraphics[width=\columnwidth]{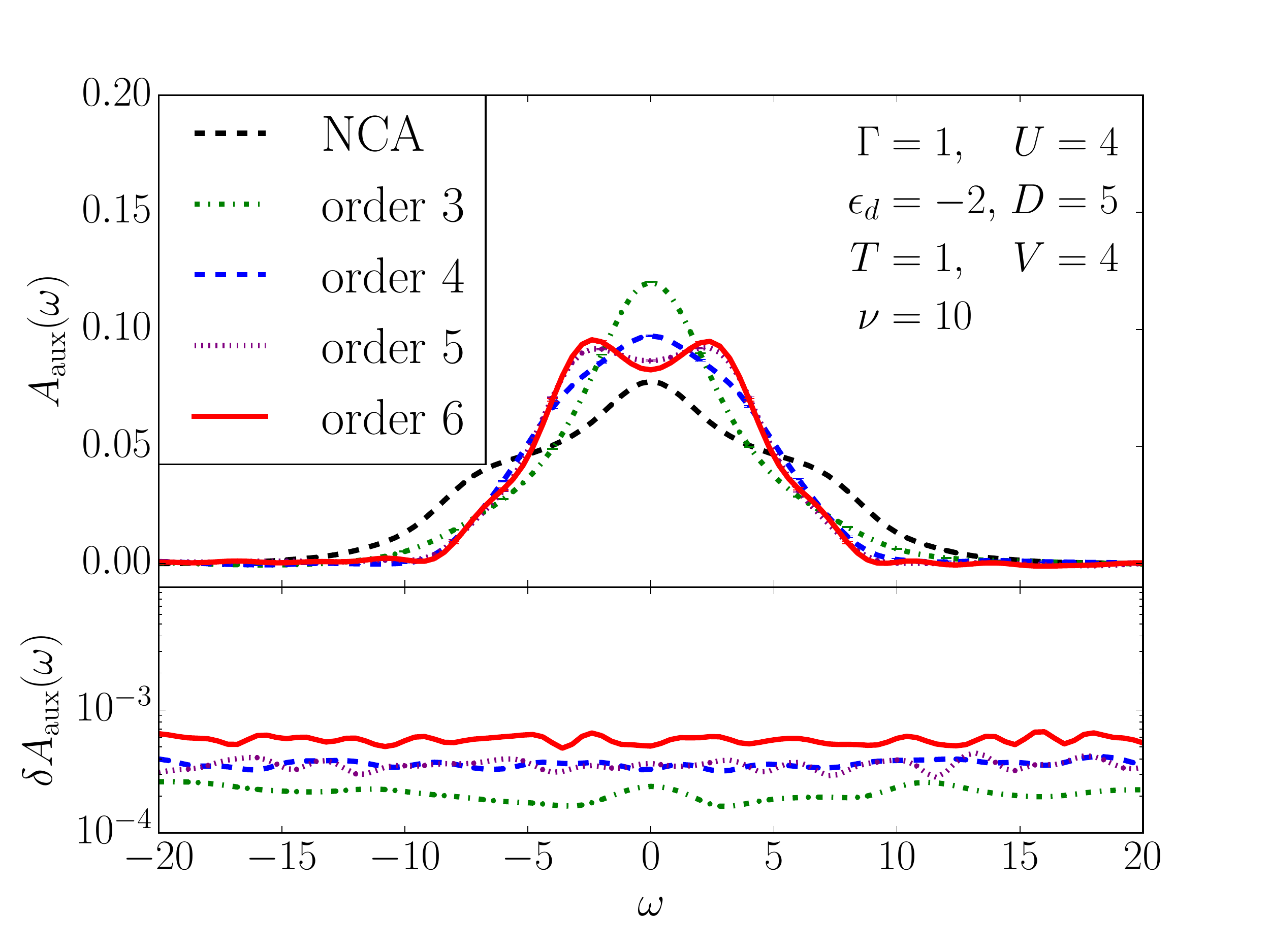}
\caption{Top panel: The (half-filling) spectrum at $t= 2.0$ after a voltage quench with $\Gamma= 1$, $U = 4$, $\epsilon_d = -2$, $D = 5$, $T = 1$ and $V = 4$. The spectral function shows the establishment of a split Kondo peak as the diagram order is increased. The data for order $6$ is identical to the data shown in Fig.~\ref{fig:auxA}. Bottom panel: Error estimate for data shown in main panel. The error remains constant as a function of frequency and increases as the maximum order is increased.}\label{fig:orderVQuench}
\end{figure}
hFig.~\ref{fig:orderVQuench} shows the convergence of the data shown at the final time $t=2.0$ in Fig.~\ref{fig:auxA} as a function of the maximum diagram order sampled. It is evident that high orders $\gtrsim 5$ are needed to accurately capture the split peak, hinting that its correct description is related to strong dot--bath entanglement. It is also evident that deviations remain between orders $5$ and $6$, indicating that even higher orders may be necessary to fully capture the physics.

\begin{figure}[]
\includegraphics[width=\columnwidth]{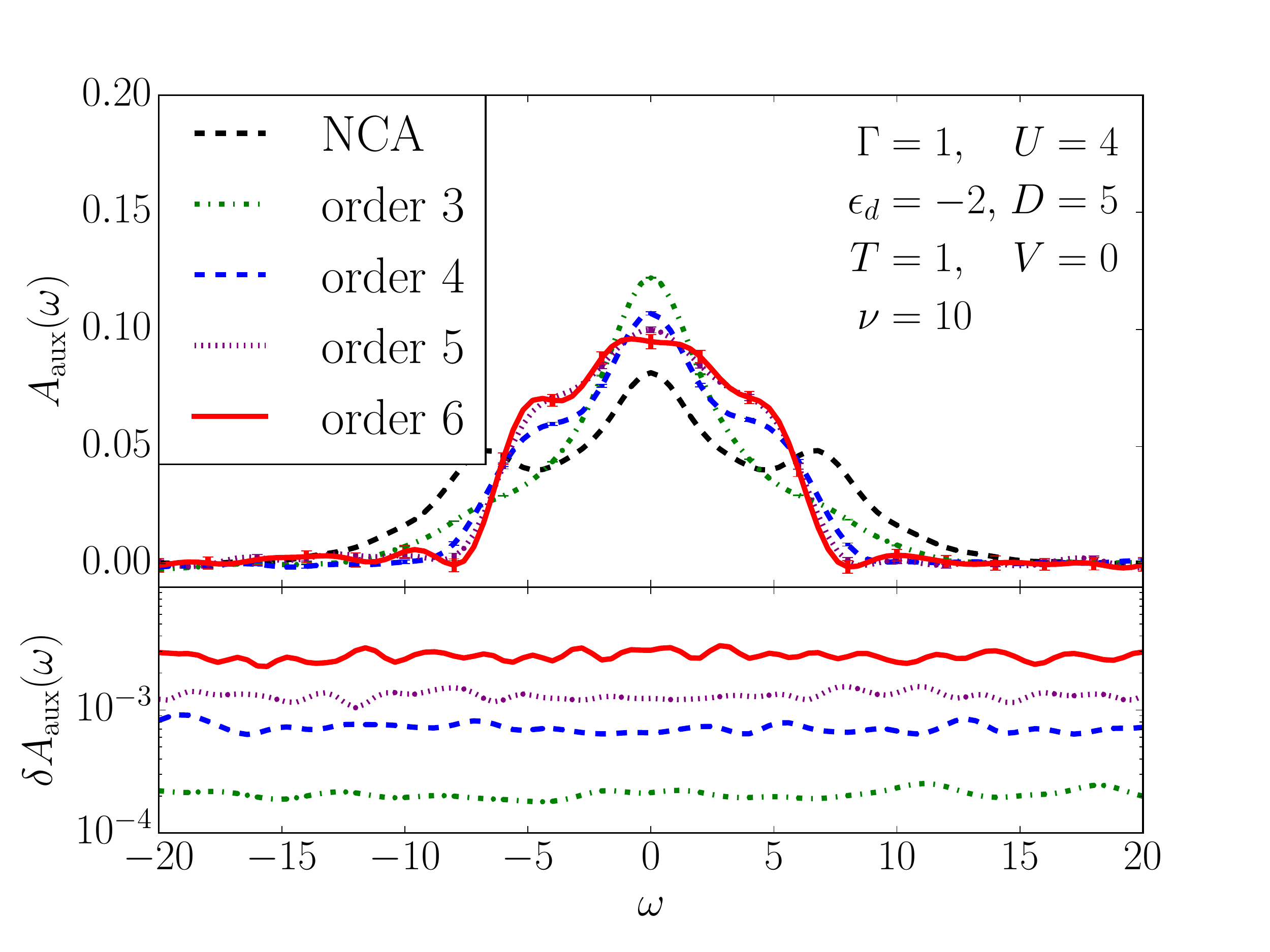}
\caption{Top panel: The (half-filling) spectrum at $t = 2.0$ with no applied voltage with $\Gamma= 1$, $U = 4$, $\epsilon_d = -2$, $D = 5$, $T = 1$ and $V = 0$. Bottom panel: error estimate for data shown in the main panel.}\label{fig:EqSpectrum}
\end{figure}
This is even more pronounced in the equilibrium case, Fig.~\ref{fig:EqSpectrum}, where contributions coming from long-lived correlations cause both an increase of the statistical errors (bottom panel) and a substantial difference order-by-order (main panel).

\begin{figure}[]
\includegraphics[width=\columnwidth]{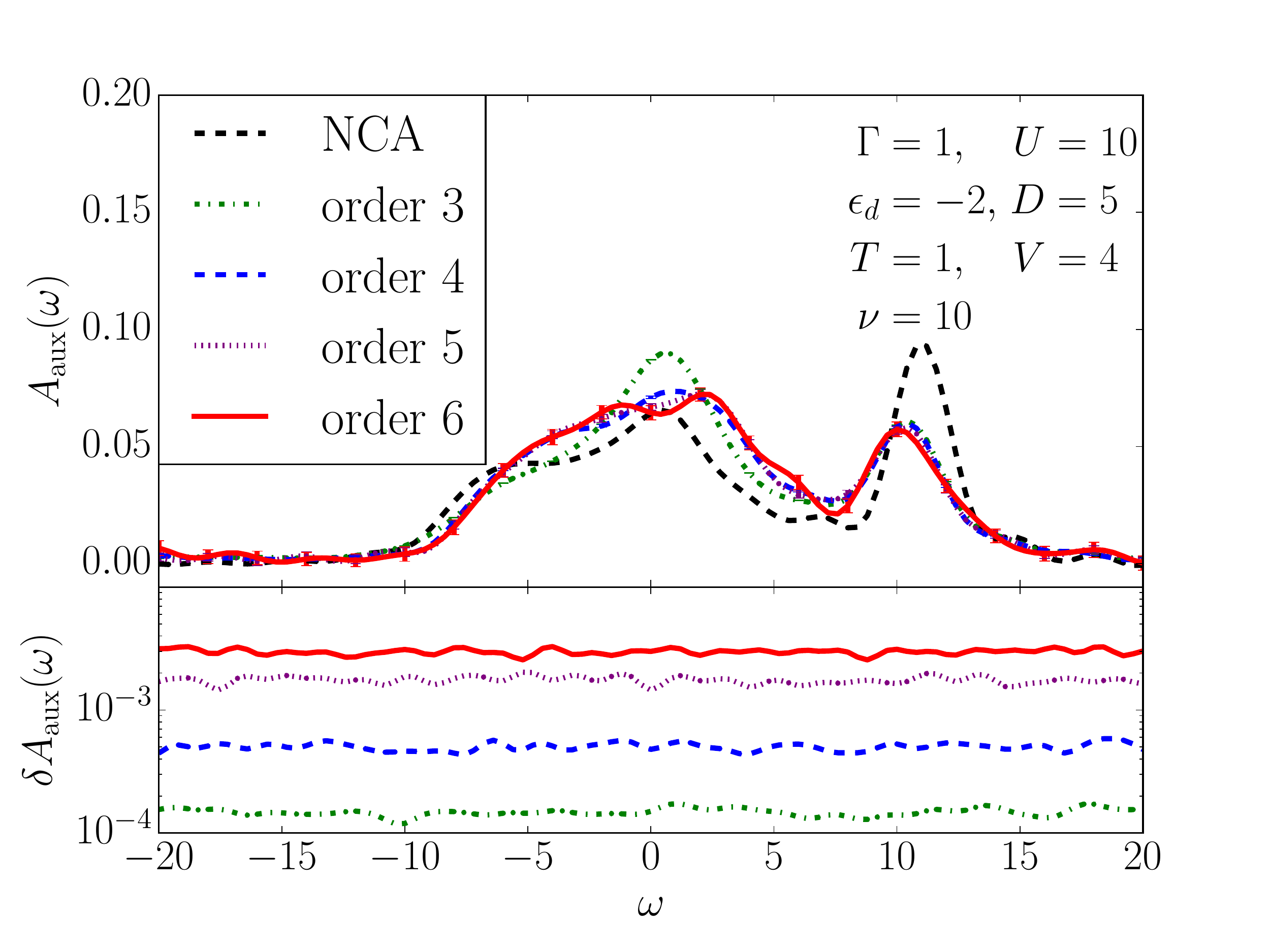}
\caption{Top panel: Spectral function away from half filling at $t = 2.0$ after a voltage quench with $\Gamma= 1$, $U = 10$, $\epsilon_d = -2$, $D = 5$, $T = 1$ and $V = 4$.  Bottom panel: error estimate for data shown in the main panel.}\label{fig:auxAawayHalf}
\end{figure}
No additional complications arise away from half filling. Fig.~\ref{fig:auxAawayHalf} shows a sample steady state spectral function of a system obtained at time $t=2.0$, away from particle-hole symmetry, after a voltage quench. The result is once again obtained with the auxiliary current setup, and is converged both in expansion order (orders $5$ and $6$ were needed) and time. While general features of the system are visible even within a low-order NCA approximation, finer details such as the precise location of the peaks or their height and width clearly require analysis with more precise methods.

\section{Conclusion}
In conclusion, we have generalized the inchworm quantum Monte Carlo method to the full forward--backward--imaginary Keldysh contour. We have also introduced a scheme to measure currents, Green's functions, and spectral functions in inchworm QMC. Our formalism for measuring these quantities is directly applicable to inchworm calculations on the forward--backward Keldysh contour, but the addition of the imaginary contour allows us to begin the simulation from a correlated equilibrium initial condition. The method is numerically exact when all diagrams to all orders are considered. It is controlled if a sequence of results truncated to gradually increasing diagram orders is considered, and in particular generates non-crossing diagrams when truncated to order one, one-crossing diagrams when truncated to order two, and two-crossing diagrams when truncated to order three. We showed that for the  applications considered in this paper, diagrams of order five to seven were sufficient to achieve convergence.

The method makes the simulation of a wide range of problem setups possible: voltage and interaction quenches out of initially thermalized states, perturbations with explicit time-dependence, long-time steady-state setups, and equilibrium problems. It can in particular be used for obtaining spectral functions in real time, eliminating the need for the numerically ill-conditioned analytical continuation procedure of imaginary time data.

Inchworm Monte Carlo overcomes the dynamical sign problem in the sense that as $t$ is increased, the effort for reaching longer times increases sub-exponentially. Unlike in the case of the forward--backward contour, we did not always observe a plateau of the error as a function of time, indicating that the scaling is generally worse than quadratic in time. Several exponential barriers remain in the system: as temperature is lowered, higher orders proliferate and the number of diagrams needed to be considered increases quickly. Similarly, a larger impurity size exponentially increases the size of the local Hilbert space and thereby the cost of simulating the system.

The results shown here illustrate that it is now possible to calculate reliable currents, Green's functions, and spectral functions for equilibrium and nonequilibrium impurity problems with general time dependence, and imply that unbiased impurity solvers, which form a fundamental component for non-equilibrium dynamical mean field theory, are now available.

\section{Acknowledgments}
This work has been supported by DOE ER 46932. This research used resources of the National Energy Research Scientific Computing Center, a DOE Office of Science User Facility supported by the Office of Science of the U.S. Department of Energy under Contract No. DE-AC02-05CH11231. GC was supported by the ISRAEL SCIENCE FOUNDATION (grant No. 1604/16).

\bibliographystyle{apsrev4-1}
\bibliography{bold_hyb}

\appendix*
\section{Inchworm equations}\label{app:eqs}
In this appendix we supplement the description of the algorithm with additional details. 

\subsection{Atomic propagators}

The bare atomic propagator for state $|\phi\rangle$ describes the evolution of the impurity from time $t_2$ to time $t_1$ at $\mathcal{V}^{\alpha}_k = 0$. It is defined as 
\begin{equation}\label{eq:bare_props}
p^{(0)}_{\phi} (t_1, t_2) = -i \xi_{\phi}^{\theta(t_1, t_2)} e^{-i \int\limits^{t_1}_{t_2} \varepsilon_\phi (t') dt' },
\end{equation}
where $t_1 \geq t_2$  are ordered on the contour and $\varepsilon_\phi$ is the energy of bare atomic state $\phi$. The factor $\xi_{\phi} = (-1)^{\langle \phi | \hat N | \phi \rangle}, \xi_\phi^2 = 1$ accounts for the bosonic statistics of the empty and doubly occupied atomic states, and $\theta(t_1, t_2) = 1 $ when $t_1 > -i \beta > t_2$, and $0$ otherwise.

The atomic propagator describes the evolution of the system, projected to a single atomic state. It is defined as
\begin{align}\label{eq:prop_define}
p_{\phi} \left(t_{1},t_{2}\right) = 
-i \xi_{\phi}^{\theta(t_1, t_2)} \mathrm{Tr}_{c} \left\{ \langle\phi | \exp\left(-i \int\limits_{t_2}^{t_1}  H(t') dt'\right) |\phi\rangle \right\}.
\end{align}
where $\mathrm{Tr}_{c}$ traces over the leads degrees of freedom.

The dot occupancy $N(t)$ can be expressed in terms of propagators 
\begin{equation}
\langle{N}\left(t\right)\rangle =\frac{i}{Z}\sum_{\phi}\langle \phi|N|\phi\rangle\xi_{\phi} p_{\phi}(t_{+},t_{-}), \label{eq:prop}
\end{equation}
where time $t_+$ and $t_-$ denote time $t$ on the `plus' and `minus' branches and the partition function $Z$ is 
\begin{equation}
Z = i\sum_{\phi}\xi_{\phi} p_{\phi}\left(t_{+},t_{-}\right). \label{eq:Z}
\end{equation}
The partition function is time-independent due to the unitarity of the evolution operator and is equivalent to the standard  definition $Z = \mathrm{Tr} \left\{ \exp \left(-\beta H\right) \right\}$ up to a multiplicative constant of the partition function of the lead electrons.

\subsection{Hybridization expansion}

As illustrated in Fig.~\ref{fig:bareprops} the value of the full atomic propagator is given by a sum over all possible connected and disconnected combinations of hybridization lines that are obtained by the expansion in the $H_T$. The equation for $p_\phi(t_1, t_2)$ reads 
\begin{multline}\label{eq:ordern}
p_\phi(t_1, t_2) = \sum_{n=0}^{\infty} i^{2n} \mathrm{Tr} \Big\{ \langle \phi | \mathcal{T_C} \prod_{j=1}^{n} \Big(\int^{t_1}_{t_2} dt'_{2j+1} \int^{t_1}_{t_2} dt'_{2j} ~ \\ \times \sum_{\sigma_j} d^\dagger_{\sigma_j}(t'_{2j+1}) c_{\sigma_j} (t'_{2j+1}) c^\dagger_{\sigma_j} (t'_{2j}) d_{\sigma_j} (t'_{2j}) \Big) | \phi \rangle \Big\}.
\end{multline}

At every expansion order $n$, $2n$ intermediate operators are introduced and denoted by the index $j$. These operators act at times within the interval $(t_2 \rightarrow t_1)$, and the integration over these times over the Keldysh contour is performed. $\sigma_j$ is the spin index of every intermediate operator. The time-ordering operator $\mathcal{T_C}$ implies the sum over all permutations of $2n$ operators, and we define every permutation as $X$. Eq.~(\ref{eq:ordern}) is then rewritten as 
\begin{equation}
p_\phi(t_1, t_2) =  \sum_{n=0}^{\infty} \sum_{C} i^{n-1}w_{\mathrm{loc}}w_{\mathrm{hyb}}, \label{eq:prop_det}
\end{equation}
where each configuration $C$ denotes the set of intermediate times $t'_1 \hdots t'_{2n}$ and permutation $X$. The sum over $C$ is the integration over all intermediate times and sum over permutations $X$. The weight of every configuration is given by
\begin{subequations}
\begin{align}
w_{\mathrm{loc}} = &(-1)^{\mathrm{sgn}(X)}\langle\phi| p^{(0)}_{\phi}(t_1,t'_{2n})\times \label{eq:det_loc_weight}
\\ \notag&  p^{(0)}_{\phi_{2n-1}}(t'_{2n},t'_{2n-1})\hdots 
 p^{(0)}_{\phi}(t'_{1},t_{2})|\phi\rangle \\
w_{\mathrm{hyb}} = & i(-1)^{k} \det \mathbf{\Delta}. \label{eq:det_hyb_weight}
\end{align}
\end{subequations}
The local weight $w_{\mathrm{loc}}$ consists of the product of bare atomic propagators and the sign of the permutation $X$. The hybridization weight $w_{\mathrm{hyb}}$ is the result of averaging of a product of $c$- operators with a non-interacting bath. This average is evaluated with the help of Wick's theorem and can be represented as a determinant of a matrix $\mathbf{\Delta}$ with columns defined by times of $c^\dagger$ operators and rows -- by times of $c$-operators. Each element of matrix $\mathbf{\Delta}$ is the hybridization function defined as
\begin{multline}
\Delta_{\sigma} \left(t, t' \right) =  \sum_k \mathcal{V}_k^2 \langle \mathcal{T}_{\mathcal{C}} c_{\alpha k\sigma} (t) c^\dagger_{\alpha k\sigma} (t') \rangle = \\
  -i \int d\omega e^{-i \omega \left(t - t' \right) } \Gamma \left( \omega \right) \left[ \theta_{\mathcal{C}} \left(t, t' \right) \pm f_{\alpha} \left( \omega \right) \right]  \label{eq:delta}
\end{multline}
where $\Gamma(\omega) = \sum_{\alpha} \Gamma^{\alpha}(\omega)$ is defined in Eq.~(\ref{eq:gamma}) and $f_{\alpha}(\omega)=1/(1+\exp(\beta(\omega-\mu_\alpha))$ is the Fermi function.

The determinant of matrix $\Delta$ is by definition 
\begin{equation}
\det \Delta = \sum_{\mathrm{all}} (-1)^{\mathrm{sgn}\{D_m\}} D_m(t'_{2n}\hdots t'_{1}).
\end{equation}
It consists of a sum over all possible diagrams $D_m$, each of which is a product of $n$ hybridization lines connecting pairs of operators with the same spin $\sigma = \uparrow$ or $\downarrow$ at times $t_1 \hdots t_{2n}$ according to a permutation $m$ of $(1,\hdots, 2n)$. These diagrams are shown in Fig.~\ref{fig:bareprops}. 

In order to maintain convention with the standard definition of the Green's function on the Keldysh contour hybridization functions $\Delta$ are time-ordered on the standard `plus' - `minus' - `imaginary' contour. The difference in time-orderings between the hybridization functions and atomic propagators results in the additional sign factor $(-1)^k$ introduced to Eq.~\ref{eq:det_hyb_weight}. Stochastic summation of  Eqs.~(\ref{eq:prop_det}) - (\ref{eq:det_hyb_weight}) constitutes the non-equilibrium CT-HYB algorithm introduced in Ref.~\onlinecite{Muhlbacher08,Werner09,Schiro09,Antipov2016}. 

\subsection{Inchworm summation}

The inchworm expansion reuses propagators obtained at shorter time intervals. Assuming that the atomic propagators are known in the interval $[t_2 \rightarrow t_{\mathrm{split}}]$, with $t_2 < t_{\mathrm{split}} < t_1$ Eqs.~(\ref{eq:det_loc_weight}),(\ref{eq:det_hyb_weight}) are written as 
\begin{subequations}
\begin{flalign}  \label{eq:inch_loc_weight}
& w_{\mathrm{loc}} = i (-1)^{\mathrm{sgn}(X)}\langle\phi| p^{(0)}_{\phi}(t_1,t'_{2n})\hdots
\\ \notag&   \hdots p^{(0)}_{\phi_{2m-1}}(t'_{2m},t_{\mathrm{split}}) p_{\phi_{2m-1}}(t_{\mathrm{split}} ,t'_{2m-1})\hdots 
p_{\phi}(t'_{1},t_{2})|\phi\rangle \\
& w_{\mathrm{hyb}} =  i(-1)^{k}\sum_{\mathrm{connected}} (-1)^{\mathrm{sgn}\{D_m\}} D_m(t'_{2n}\hdots t'_{1}). \label{eq:inch_hyb_weight}
\end{flalign}
\end{subequations}

Here only the subset of ``connected'' diagrams is used in evaluation of the hybridization weight. The diagram selection rules are described in Sec.~\ref{sec:inchworm} and illustrated in Fig.~\ref{fig:inchprops}. 

\subsection{Green's function diagrams}

Similarly to the propagators, the hybridization expansion for the Green's function $G(t_{\mathrm{split}}, t_2) = -i \langle \mathcal{T_C} d(t_{\mathrm{split}}) d^\dagger(t_2) \rangle $ can be written as a product of local and hybridization weights. The resulting expression reads
\begin{equation}
Z G(t_{\mathrm{split}}, t_2) = -\sum_{n=0}^{\infty}\sum_{C} i^{n}w^G_{\mathrm{loc}}w^G_{\mathrm{hyb}},
\end{equation}
where an additional auxiliary hybridization line $\Delta^{G}(t_{\mathrm{split}}, t) = -i$ is added to the configuration that now consists of $2n+2$ points. The corresponding weights are 
\begin{subequations}
\begin{flalign}  \label{eq:gf_loc_weight}
& w^G_{\mathrm{loc}} = i (-1)^{\mathrm{sgn}(X)}\langle\phi| p_{\phi}(t_1,t'_{2n})\hdots
\\ \notag&   \hdots p_{\phi_{2m}}(t'_{2m}, t_{\mathrm{split}}) p_{\phi_{2m-1}}(t_{\mathrm{split}} ,t'_{2m-1})\hdots 
 p_{\phi}(t'_{1},t_{2})|\phi\rangle \\
& w^G_{\mathrm{hyb}} =  i(-1)^{k}\sum_{\mathrm{all~crossing}} (-1)^{\mathrm{sgn}\{D_m\}} D_m(t'_{2n+2}\hdots t'_{1}). \label{eq:gf_hyb_weight}
\end{flalign}
\end{subequations}
Time $t_2$ here belongs to the `minus' branch of the contour, and $t_1$ is equal to $t_2$, but belongs to the `plus' branch. The choice of diagrams that only cross the auxiliary hybridization line (denoted as `all crossing') is discussed in Sec.~\ref{sec:gf} and illustrated in Fig.~(\ref{fig:inchgf}). 

\subsection{Spectral function from auxiliary leads} \label{subsec:aux}
A convenient method for extracting the spectral functions $A(\omega)$
from non-equilibrium Green's functions $G \left(t, t'\right)$ comes from
considering the current through two auxiliary leads which are weakly coupled
to the system only at a predefined frequency $\omega' \ [\Gamma(\omega) =
\eta\delta(\omega - \omega')]$, where one lead is taken to be full  and one to be empty.\cite{cohen_greens_2014}
Using Eq.~\ref{eq:delta}
this leads to hybridization functions given by
\begin{equation}
\Delta_{\rm aux}(t, t') = -i \eta e^{-i\omega'(t - t')}[\theta_\mathcal{C}(t, t') - f_i]
\end{equation}
where 
\begin{equation}
f_i =  
\begin{cases}
0: i = 0 \ (\mathrm{empty})  \\
1: i = 1 \ (\mathrm{full}) \\
\end{cases}
\end{equation}
and $\eta$ is small.
Using equation 
\begin{equation}\label{eq:current}
\left\langle I_\alpha(t)\right\rangle = 2\Re\left(\sum_{\sigma}\int_{\mathcal{C}}dt' G_\sigma \left( t', t \right)  \times\Delta_{\alpha\sigma}(t,t')\right)
\end{equation}
to calculate the currents $I_A^\text{e}
\left(\omega, t\right)$, $I_A^\text{f} \left(\omega, t\right)$ through the
empty and full auxiliary leads, we define the object
\begin{equation}
A_\text{aux} \left(\omega, t\right) = \lim_{\eta \rightarrow 0} - \frac{2h}{e \pi \eta} \left[ I_A^\text{f} \left(\omega, t\right) - I_A^\text{e} \left(\omega, t\right) \right]
\end{equation}
which reproduces the spectral function $A(\omega, t)$ in steady state
\cite{cohen_greens_2014-1,cohen_greens_2014}.  Here, we perform this process as a
post processing step on our non-equilibrium Green's functions so that the auxiliary
leads are not included in our simulations.
In our experience this way of obtaining the spectral information provides results that are more stable than an explicit Fourier transform of the time-dependent Green's function.

\end{document}